\documentclass[epj]{svjour}
\usepackage{graphics}
\begin{document}
\title{On the stability of hole crystals in layered cuprates.}
\author{G. Rastelli\inst{1} \and S. Fratini\inst{2} \and P. Qu\'emerais\inst{2}
}                     
%
%
\institute{Istituto Nazionale di Fisica della Materia and
Dipartimento di Fisica,
Universit\`a dell'Aquila,\\ via Vetoio, I-67010 Coppito-L'Aquila, Italy 
\and  Laboratoire d'Etudes des Propri\'et\'es Electroniques des
   Solides, CNRS \\
BP166 - 25, Avenue des Martyrs, F-38042 Grenoble
   Cedex 9, France
}
\date{Received: date / Revised version: date}
%
\abstract{
Recent STM measurements have revealed the existence of periodic charge
modulations at the surface of
certain cuprate superconductors. 
Here we show that the observed  patterns are 
compatible with  the formation of a three-dimensional  crystal
of doped holes,
with space correlations
extending between different Cu-O layers.
This puts severe constraints on the dynamical stability
of the crystallised hole structure, resulting in a close relationship
between  the periodicity of the electronic modulation and the interlayer distance.
} 
\maketitle

Since the discovery of high temperature superconductivity, 
several microscopic models have been proposed in order to explain
the complex phase diagram of the cuprates. In particular, 
the presence of an antiferromagnetic phase in the parent
compounds has led to a huge theoretical effort on models with local
electronic interactions, such as the Hubbard or  t-J models.
It is seldom realised that, when few carriers are added to an
insulating system by doping, the long range part of the Coulomb repulsion 
is not screened, and can not be neglected \textit{a priori}. 

Interest in the long range interactions has been recently revived by 
the observation, by scanning tunneling microscopy, of periodic
modulations of the electronic density of states
at the surface of the cuprate compounds
Bi$_2$Sr$_2$CaCu$_2$O$_{8+\delta}$  (Bi-2212)
\cite{Howald-unpub-02,Hoffman-sci295-02,Hoffman-sci297-02,Howald-PRB03,McElroy-nat03,Vershinin-04,McElroynodal-04,Fang-04}
and Ca$_{2-x}$Na$_x$CuO$_2$Cl$_2$ (Na-CCOC). \cite{Hanaguri-04}
Although some of the features have been interpreted
in terms of quasiparticle interference
effects, the emergence of  modulations with a dispersionless
ordering vector $q\simeq 2\pi/4a_0$ directed along the Cu-O
bonds, strongly points to the existence of an underlying charge
order, characteristic of the less conducting (pseudogap)
regions. Among  other possibilities, 
the latter could be ascribed   
to the formation of a 
Wigner crystal \cite{Wigner} of holes 
\cite{Hanaguri-04,Remova,queque,frat-redshift,Kim,FuDavis-04}
--- an insulating ordered state arising in
electronic systems at low density, when the long-range Coulomb interactions
are dominant --- or to the ordering of hole pairs. 
\cite{Zhang,Vojta,Tesanovic,Anderson}

The aim of this work is to determine 
if the concept of hole crystallisation is compatible
with the experimental observations in the cuprates. 
It is clear that a complete microscopic treatment of the
problem should consider, in
addition to the aforementioned long range
Coulomb repulsion, also the interaction of the doped  
holes with all the 
degrees of freedom  of the host material, including the $1-x$
localised electrons in the Mott insulator, the ions of the host
lattice, chemical impurities, etc... 
Instead of attempting this formidable task, or 
choosing to rely on a definite microscopic model, 
we shall approach the problem starting from the experimental observation 
that the holes \textit{are} localised in an ordered pattern, and 
describe the system
in the framework of a phenomenological Lorentz model. 
As we will show, a great insight into the problem can be gained already 
at this phenomenological level,
leading to precise constraints on the admissible shape and periodicity
of the charge ordering patterns.

\begin{figure}
\centerline{\resizebox{0.7 \columnwidth}{!}{\rotatebox{0}{\includegraphics{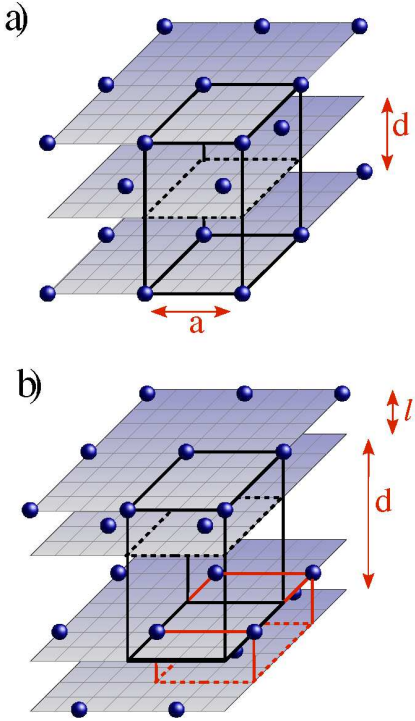}}}}
\caption[]{The anisotropic crystal structures considered
  in the text:  a) single layer compounds ($d$ is the interlayer spacing, 
  $a$ is the hole-hole distance within a
  layer); b) bilayer compounds ($l$ is the distance between 
  neighbouring layers).}
\label{fig:structures}
\end{figure}

Let us consider the holes to be located at the sites $\bf{R}_i$
of a  \textit{three-dimensional} Bravais lattice. 
This is a sensible assumption
because, despite the structural
constraints in the cuprates, which confine the carrier motion into
Cu-O layers,  the long range  Coulomb 
interactions are isotropic, so that  
the ordering within a Cu-O layer is certainly affected by the
positions of the holes on different layers. 
Allowing for small (inplane) displacements $\bf{R}_i\to\bf{R}_i+
\bf{u}_i$  around 
the equilibrium positions leads to the 
following quadratic Hamiltonian:
\begin{equation}
  \label{eq:H}
  H_I= \sum_i H_i + 
\frac{1}{2\epsilon_\infty}\sum_{i\neq j}  u_i^\alpha
\mathcal{I}^{\alpha \beta}_{ij}
  u_j^\beta + \sum_i \frac{1}{2}m\omega_0^2 u_i^2.
\end{equation}
The first term on the right represents all the \textit{static} 
long-range Coulomb interaction effects, including  
the Madelung energy of the Bravais lattice, plus the restoring
potentials acting at each $\bf{R}_i$, due to
the repulsion of the other holes localised at $\bf{R}_j\neq
\bf{R}_i$.  
The second term
is the usual  dipole-dipole
interaction arising between particle oscillations. The indices
$\alpha,\beta=x,y$ are summed,  and
$\mathcal{I}^{\alpha \beta}_{ij}=\frac{|R_{ij}|^2\delta^{\alpha\beta}
  -3R_{ij}^\alpha R_{ij}^\beta }{|R_{ij}|^5}$. It should be noted
that  the  terms $H_i$ are screened by the  
static dielectric constant $\epsilon_s$ of the host medium, while 
the dipolar interactions between fast hole oscillations
are ruled by 
the high frequency  dielectric constant  $\epsilon_\infty$. 
This discrepancy can have important consequences in the cuprates,
where $\epsilon_s$ is substantially larger than $\epsilon_\infty$
owing to the ionic polarisability.
The last term represents 
the (unknown) potential which, within our phenomenological approach, 
accounts for the influence of the host material on the dynamics of the
individual holes (interaction with the magnetic, ionic degrees
of freedom...).

Let us first consider a ``pure'' Wigner crystal, which is obtained
by setting $\omega_0=0$ in the above Hamiltonian.
\cite{Bagchi}
Among the different three-dimensional Bravais lattices compatible with
the structural 
anisotropy imposed by the Cu-O layers, we shall deliberately restrict
ourselves to 
the cases which present a \textit{square} charge ordering within the
planes, as 
observed in experiments. 
Within this subclass, it can be demonstrated that 
the body centered tetragonal (BCT), illustrated in
Fig. 1.a., has the lowest Madelung energy. 
\begin{figure}
\centerline{\resizebox{.7 \columnwidth}{!}{\rotatebox{0}{\includegraphics{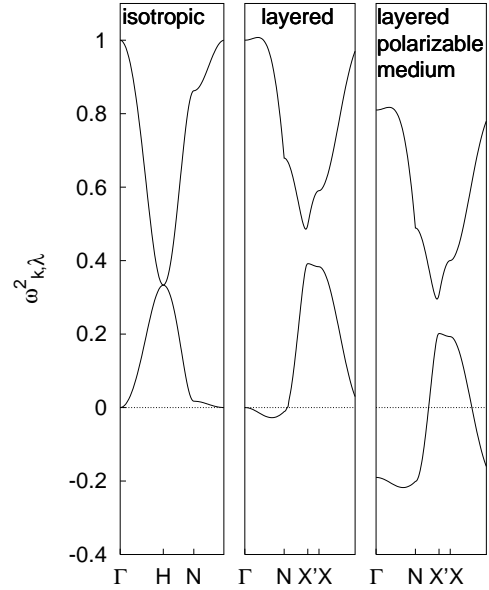}}}}
\caption[]{Spectrum of the collective excitations 
of different Wigner crystals, for wavevectors $k$
within the plane.
Left panel: the usual body centered cubic (BCC)
Wigner lattice ($\gamma=1$).
Center panel: anisotropic BCT lattice with $\gamma=0.5$, which
is unstable against shear, as 
signaled by the existence of purely imaginary solutions 
with $\omega_{{\bf k} \lambda}^2<0$. 
Right panel:  the
same anisotropic lattice embedded in a polarisable medium with 
$\eta=\epsilon_\infty/\epsilon_s=0.5$. 
The eigenvalues are  expressed
in units of $\omega_p^2=4\pi
ne^2/m\epsilon_\infty$, $m$ being the hole mass, and $n$ the
hole density.  
 The symmetry points indicated in the graphs are defined as follows, 
in units of $\pi/a$. Left panel: $\Gamma=(0,0)$, H$=(2,0)$,
N$=(1,1)$. Centre and right panel: $\Gamma=(0,0)$, N$=(1+\gamma^2,0)$
X$^\prime=(1+\gamma^2,1-\gamma^2)$, X$=(1,1)$.
}
\label{fig:eigenfrequencies}
\end{figure}

To determine if  such anisotropic  Wigner crystal 
is stable, we now evaluate the frequencies $\omega_{{\bf k}\lambda}$
($\lambda=$ branch index) of the corresponding collective charge density oscillations. 
The calculation is performed using standard Ewald summation techniques, 
at different values of the anisotropy ratio
$\gamma=a/2d$ ($a$ being the inplane interparticle spacing, determined
by the hole concentration, and $d$ the
interlayer distance). 
The excitation spectra
are reported in figure 2.
In the isotropic case  
($\gamma=1$, left panel), we recover the 
body centered cubic (BCC) structure, which is 
known to be mechanically stable. 
Generic BCT crystals are also stable for weak anisotropy ratios within the range  $0.66<\gamma<1.07$.
Note that this identifies an interval of hole
concentrations, where a square ordering in the planes arises
naturally
due to the three-dimensional character of the Coulomb interactions.

For anisotropy ratios outside the range  $0.66<\gamma<1.07$, the emergence of
imaginary eigenfrequencies ($\omega^2_{{\bf k}\lambda}<0$,
see fig.2, central panel)    signals
that the  BCT structure considered here is, in principle, mechanically
unstable, so that the preferred Wigner crystal structure has a
different symmetry.
However, energetic differences
between competing structures are expected to be very small, since
Madelung energies are essentially determined by long range
effects. Therefore, it is likely that the observed hole distributions are
influenced by other microscopic mechanisms, such as
the commensurability with the Cu-O host lattice, which justifies our
restriction to square planar orderings.
\cite{note2}

Let us now consider a   crystal of holes  embedded in a polarisable
medium. In this case, 
an additional source of instability appears, causing an overall downward
shift of the excitation spectrum (fig. 2, right panel).
This occurs because, owing to the \textit{ionic polarisability}
(i.e. as soon as $\eta=\epsilon_\infty/\epsilon_s<1$), the static restoring
potentials that localise the holes, and their mutual dipolar interactions, 
responsible for the dispersion of the collective frequencies, 
are screened by different dielectric constants.
We conclude that some additional localising mechanism is required to
stabilise a hole crystal in polarisable media such as the cuprates,
and turn our attention to the full Lorentz model of eq. (1).
The collective
frequencies $\Omega_{{\bf k}\lambda}$ in this case can 
be derived straightforwardly from
the solution of  the pure Coulomb problem as: \cite{Bagchi}
\begin{equation}
  \label{eq:freqs}
  \Omega_{{\bf k}\lambda}^2=\omega_{{\bf  k}\lambda}^2+\omega_0^2.
\end{equation}
Mechanical stability requires that the 
eigenfrequencies $\Omega_{{\bf k}\lambda}$ 
are all real. This, according to equation (2), 
sets a lower bound on the 
energy scale of the localising potential: $\omega_0^2\geq \omega^2_{min}$,
where $\omega^2_{min}$ is defined as 
the modulus of the most negative eigenvalue 
$\omega_{{\bf k}\lambda}^2$ over the
Brillouin zone. 
\begin{figure}
\centerline{\resizebox{.85 \columnwidth}{!}{\rotatebox{-90}{\includegraphics{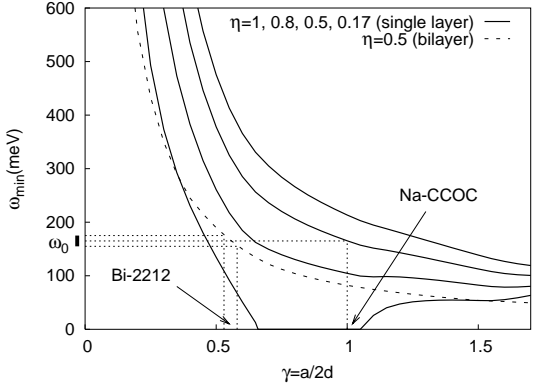}}}}
\caption[]{
The characteristic
energy scale of the  trapping potential 
necessary to stabilise an anisotropic hole crystal, as a function of the 
anisotropy ratio  $\gamma$.
Full lines corresond to the single layer host structure of Na-CCOC,
with $d=7.75 \mbox{\AA}$ and $\epsilon_\infty=4.5$,
at different values of the polarisability ratio (from left to right,
$\eta=1,0.8,0.5,0.17$). The dashed line is for the bilayer
structure of Bi-2212, with $\epsilon_\infty=4.9$ and $\eta=0.5$.        
The system becomes increasingly unstable upon 
increasing the hole concentration (i.e. reducing $\gamma$).   
The different periodicities observed in Na-CCOC 
and in Bi-2212 (indicated by arrows) correspond to a
common localisation energy $\sim 0.16 eV$, characteristic of the host
cuprate materials.}
\label{fig:stability}
\end{figure}
This quantity  is reported in figure 3 as a function of $\gamma$, for
different values  of the  polarisability ratio 
$\eta$. 
We see that it is low and flat at
large $\gamma$ 
(i.e. at low hole concentrations), 
but it  rises sharply for $\gamma \approx < 0.5$,
indicating that charge patterns whose inplane periodicity is
much shorter than the interlayer distance
are very unlikely to occur, since these can only be stabilised by
invoking an additional localisation
mechanism with a very large energy scale. 
This general observation establishes
a direct link  between  the admissible periodicity 
of the charge modulations ---an electronic property--- and   the
distance between Cu-O layers ---a structural constraint. 
To be more specific, 
using the appropriate values $d=7.75 \mbox{\AA}$, $a_0=3.85\mbox{\AA}$, $\epsilon_\infty=4.5$ and
$\eta\approx0.5$ \cite{Zibold-96} for Na-CCOC, 
it can be read directly from figure 2 that the observed periodicity
$a=4a_0$ ($\gamma=1$) 
\textit{implies} that the characteristic energy scale 
involved in the localisation of the holes is (at least)
$\omega_0\approx 0.16 eV$.

If the proposed picture is correct, it should apply to other
cuprates, such as the bismuth based
Bi$_2$Sr$_2$CaCu$_2$O$_{8+\delta}$. 
These compounds present an additional constraint related to the
bilayer structure,
for which the corresponding lowest energy configuration 
is illustrated in fig 1.b. As in the single
layer case, the calculated  excitation spectrum shows that 
a pure Wigner crystal of holes is unstable due to the ionic polarisability,
and an additional source of localisation
must be considered. Repeating the same arguments as in Na-CCOC, with 
parameters $d=15.5 \mbox{\AA}$, $l=3\mbox{\AA}$ and 
dielectric constants $\epsilon_\infty=4.9$,
$\eta\approx 0.5$,\cite{Kovaleva-04}  we read from figure 3 (dashed line) that
the observed periodicities $a/a_0=4.3-4.7$
\cite{Howald-unpub-02,Howald-PRB03,Vershinin-04,McElroynodal-04,Fang-04}
($\gamma=0.53-0.58$)
are stabilised by  taking $\omega_0=0.15-0.17 eV$, which is remarkably 
close to the value obtained in Na-CCOC. 
This result strongly points to the existence of a
\textit{common} localisation mechanism, which could be at the origin
of the different periodicities observed  in 
different compounds.
Following the same steps
in the case of 
the single layer compound Bi$_2$Sr$_2$CuO$_6$, which has a shorter
interlayer distance than  Na-CCOC ($d=12.3\mbox{\AA}$),
we predict a  crystal periodicity  $a/a_0\approx 3.4$ 
that would be interesting to verify experimentally.

Before enquiring about
its microscopic foundations,
let us analyse the consequences of the present scenario.
First of all, since $\omega_0$ reflects the interaction of individual
holes with the host material, 
in principle it should not depend much on the doping level. Here  we take
it as a constant, characteristic of
the parent cuprate compounds.
According to equation (2), 
in the limit of  vanishing hole density, where all the  $\omega_{{\bf
    k}\lambda}\to 0$, the excitation spectrum will  be dominated by $\omega_0$.
Upon increasing the hole concentration, however, the
collective frequencies spread away from
$\omega_0$ due to the term $\omega_{{\bf k}\lambda}^2$, which  admits
both positive and negative values. As a consequence,
the longitudinal charge oscillations get hardened upon doping, while 
transverse oscillations are progressively softened, until a given
eigenfrequency $\Omega_{{\bf k}={\bf k}_c}$ vanishes, leading to a
\textit{polarisation  catastrophe} at a critical concentration $x_c$.
\cite{frat-redshift,Bagchi,Herzfeld,frat-dipolar} 
The predicted softening of long wavelength transverse oscillations,
\cite{frat-redshift,frat-dipolar} which  is entirely due to
the long-range polarisability of the medium,  
has been experimentally  observed in
systematic  studies of  the optical conductivity spectra of 
both electron- and hole-doped cuprates in the underdoped region.
\cite{lupiNCCO,lupiLSCO} 
Similar signatures of the polarisation catastrophe can be expected
in spectroscopic ellipsometry and electron energy loss spectroscopy.

In the foregoing discussion,
we have assumed that  all of the doped holes are 
crystallised, so that  the system is
insulating for $x<x_c$. In this case, 
the periodicity of the modulation varies continuously 
with $x$ until it reaches its
limiting  value $a/a_0=(n_l/x_c)^{1/2}$ at the critical concentration 
($n_l=1,2$ for single and bilayer host structures).
\cite{note3} 
When further holes are doped into the system beyond  $x_c$, 
these can not be accomodated in the crystallised state, 
which is at the border of an instability. 
If, as indicated by experiments, 
the hole crystal survives for $x>x_c$,
one possibility is that  
the excess holes settle in the interstitials of the
main ordering pattern, creating a superimposed modulation, as
observed in the Na-CCOC samples.
\cite{Hanaguri-04} 
 Another possibility is that the system phase separates into insulating 
(crystallised) and conducting regions, as seems to be the case in
Bi-2212.\cite{McElroynodal-04}
The question of crystal stability in the presence of additional mobile charges,
and the consequent screening of the long range interactions, remains open.

Based on the emerging scenario, we can now speculate
on the microscopic mechanism underlying the localisation of the holes.
First of all, the energy scale $\omega_0$ identified above is definitely
too large to be imputed exclusively to the binding potentials of
chemical impurities, or to the pinning by commensurability effects.
\cite{fnote}
On the other hand, 
the interaction with the antiferromagnetic background, 
which is ruled by typical exchange energies $J\sim 0.1 eV$, 
would constitute a viable possibility.
However, in strongly polarisable materials, the excess charges added
by doping are expected to form dielectric polarons. Indeed,
it can be recognised 
that the value of $\omega_0$ 
derived in this work coincides with the locus of 
a well defined absorption band, ubiquitous in the
infrared optical spectra $\sigma(\omega)$ of  strongly
underdoped cuprates, which is generally ascribed to the formation of
polarons,\cite{Kastner-review} suggesting that the additional 
mechanism required to stabilise the hole crystal 
is the \textit{polaron self-trapping potential}.\cite{Frohlich}
The Wigner crystallization of polarons in the insulating phase of 
the cuprates has been  proposed independently  
by Remova et al.,\cite{Remova} and 
Qu\'emerais. \cite{queque}  
The quantum melting of such polaron crystal has been studied 
in refs. \cite{frat-redshift,frat-dipolar,frat-meanfield,rastelli}
yielding similar conclusions as in the present paper.

Our analysis demonstrates that the periodic modulations observed
at the surface of the cuprates 
are compatible with the crystallisation
of holes, arising due to the combined effects of the long range Coulomb
interactions, and of some additional localising phenomenon, whose 
characteristic energy scale is $\omega_0\sim 0.16 eV$.
Although several microscopic mechanisms can 
be involved,
polaron formation appears as a good candidate to stabilise a hole crystal 
in the cuprates.

\bigskip

The authors thank D. Mayou and B.K. Chakraverty for useful discussions.
G.R. acknowledges kind hospitality at LEPES, Grenoble.


\begin{thebibliography}{99}
\bibitem{note3} Assuming for simplicity that the Lorentz local field
factor of the crystallised structure at the concentrations of interest
does not deviate significantly from the isotropic case, 
we can write the following approximate expression for the 
critical concentration:
\[
  x_c=\frac{3 \epsilon_\infty}{16 \pi(1-\eta)}\left(\frac{\hbar
    \omega_0}{Ryd}\right)^2 \frac{da_0^2}{a_B^3}
\]
where $a_B$ is the Bohr radius and $Ryd=13.6 eV$. Taking 
$\epsilon_\infty\approx 4.5$, $\eta\approx 0.5$ and $\omega_0\approx
0.16eV$ yields $x_c=3.9\times 10^{-3} (d/a_B)$ which gives 
$x_c\approx 0.05-0.11$ for several cuprate crystal structures,
with corresponding modulations $a/a_0\approx 3.5-6$.
\end{thebibliography}
\end{document}